# LeanAI: A method for AEC practitioners to effectively plan AI implementations

Ashwin Agrawal[1], Vishal Singh[2], and Martin Fischer[3]

[1]Department of Civil and Environmental Engineering, Stanford University, Stanford, CA, USA
[2]Centre of Product Design and Manufacturing, Indian Institute of Science, Bangalore, India
[3]Department of Civil and Environmental Engineering, Stanford University, Stanford, CA, USA

ashwin15@stanford.edu, singhv@iisc.ac.in, fischer@stanford.edu

**Abstract –**

Recent developments in Artificial Intelligence (AI) provide unprecedented automation opportunities in the Architecture, Engineering, and Construction (AEC) industry. However, despite the enthusiasm regarding the use of AI, 85% of current big data projects fail. One of the main reasons for AI project failures in the AEC industry is the disconnect between those who plan or decide to use AI and those who implement it. AEC practitioners often lack a clear understanding of the capabilities and limitations of AI, leading to a failure to distinguish between what AI should solve, what it can solve, and what it will solve, treating these categories as if they are interchangeable. This lack of understanding results in the disconnect between AI planning and implementation because the planning is based on a vision of what AI should solve without considering if it can or will solve it. To address this challenge, this work introduces the LeanAI method. The method has been developed using data from several ongoing longitudinal studies analyzing AI implementations in the AEC industry, which involved 50+ hours of interview data. The LeanAI method delineates what AI should solve, what it can solve, and what it will solve, forcing practitioners to clearly articulate these components early in the planning process itself by involving the relevant stakeholders. By utilizing the method, practitioners can effectively plan AI implementations, thus increasing the likelihood of success and ultimately speeding up the adoption of AI. A case example illustrates the usefulness of the method.

**Keywords –**

Artificial Intelligence; Machine Learning; Technology management; Lean methodology; Last planner system; AI strategy; AI adoption; Technology adoption; Digital strategy

## 1   Introduction

Recent developments in Artificial Intelligence (AI) provide unprecedented automation opportunities in the Architecture, Engineering, and Construction (AEC) industry [1]. It offers the potential to significantly increase productivity, add enormous value, enhance competitive advantage, and increase economic welfare [2], [3]. This immense potential of AI has prompted many companies to invest in developing and adapting AI capabilities. In fact, Gartner expects 70% of organizations, including AEC, to invest and operationalize some form of AI capability by 2025 [4].

However, despite the enthusiasm regarding the use of AI, 85% of current big data projects fail [5]. In a recent survey by Boston Consulting Group and MIT, seven out of ten companies reported little or no impact of AI [6]. These constant failures result in wastage of time, money, and effort. It also risks AI adoption in the AEC industry since practitioners might reject AI as a hype [7].

One of the main reasons for AI project failures in the AEC industry is the disconnect between those who plan or decide to use AI and those who implement it [8]–[11]. [12], [13] notes that the process of implementing AI projects in engineering firms remains less understood. [11] reports from a survey of 3,000 business executives that there is a big disconnect between the ambitions and implementation of AI, and more than 60% firms do not have a concrete AI implementation strategy. Therefore, research addressing the disconnect between planning and implementation is crucial. Such research can provide valuable insights and lessons for organizations just beginning their AI journey and can help them navigate the complexities of implementation and increase their chances of success [14].

In this work, we introduce the LeanAI method to alleviate the disconnect between AI planning and implementation in the AEC industry. The LeanAI method delineates what AI should solve, what it can solve, and what it will solve, forcing practitioners to clearly articulate and align these components early in the planning process itself by involving the relevant stakeholders. Based on our case studies data and practitioner feedback, it has been observed that the LeanAI method is highly valuable in facilitating better planning for AI implementations in the AEC industry (see Section 5 for details).

In the next section, the paper provides details about the research method used for developing the LeanAI method. This is followed by a brief background about the use of AI in AEC in Section 3, introducing the LeanAI method, detailing the steps on how it can be used and showcasing its usage in the context of an example in Section 4, and providing evidence for the usefulness of the method in Section 5. The paper is concluded with the discussions of the findings and their implications for the AEC industry in Section 6.

## 2 Research Method

The LeanAI method is built using an ethnographic-action research methodology [15], where the researchers helped AEC practitioners implement AI projects while simultaneously observing the challenges in AI implementation and formulating solutions. Ethnographic-action research has been popular for investigating the use of technology in construction projects [16], [17] because it allows the researchers to holistically understand the local context of AEC practitioners while building the solution, thereby increasing the likelihood of a solution that works in practice [18].

**LeanAI-method development:** The LeanAI method is built using data from two sources:

(1) An ongoing longitudinal study, which has investigated four case studies since 2018, has collected 50+ hours of interview data and conducted direct observations of AEC practitioners. The study focuses on Digital Twin and AI implementations that involve Machine Learning, Deep Learning, Computer Vision, and Natural Language Processing. This information has been documented in sources such as [8]-[10], [19], and [20].

(2) 12 graduate-level student case studies, each lasting for three months, involving AI implementations with industry partners in the USA, as part of a project-based class at Stanford University.

Each of the sixteen case studies involved an AEC company that wanted to implement some form of AI algorithm. The nature of the cases varied, with some being more exploratory in nature and others being very specific about the algorithm they wanted to implement.

The studies lasted an average of 3 to 6 months and involved three phases: ideation, AI algorithm development, and roadmap creation for future deployment. The authors were directly involved in at least one of these phases for each case study. Ethnographic techniques were used to collect data for all sixteen case studies.

Using above observations and hands-on experience of building AI projects with AEC professionals, the authors identified key components crucial to addressing the disconnect between AI planning and implementation. These included clearly defining the business needs the AI solution will address, outlining a specific problem statement for the AI to solve, identifying the required data and AI methods, and establishing metrics to measure the AI's performance. As the work progressed, the authors found that the Last Planner System [21] was effective in organizing these components in a simple, logical, and relatable structure for AEC practitioners, resulting in the development of the LeanAI method.

While the case studies demonstrated encouraging outcomes, it was crucial to conduct an independent validation of the efficacy of the LeanAI approach to ensure its robustness. This was necessary because the method was developed based on data from the case studies, and assessing its usefulness solely on those studies could introduce bias. Therefore, we conducted a workshop with practitioners as described below.

**LeanAI-method's usefulness validation:** Thirty practitioners from different companies used the LeanAI method to plan an upcoming/ongoing AI implementation in their companies. Their feedback about using the LeanAI method to plan their AI implementations was recorded and the results have been described in Section 5.

## 3 Background on Artificial Intelligence and its usage in AEC industry

AI is a rapidly growing field that encompasses a wide range of subdisciplines and application areas. Broadly, AI helps computers perceive, represent, reason, solve, and plan in an intelligent and adaptive manner, making them capable of dealing with complicated and ill-defined problems which were long thought to be the exclusive domain of humans [22].

Machine Learning (ML), a key area in AI research, involves the development of algorithms that allow computers to learn from data (and labels) to make predictions, decisions, and perform tasks for which they were not explicitly programmed. For example, Natural Language Processing (NLP), a sub-part of ML, involves development of algorithms that can understand, interpret, and respond to human language. NLP is used in a wide range of tasks such as language translation and text summarization for which writing an exhaustive set of rules to explicitly program the computer is nearly impossible. Another popular field within ML is Computer Vision. This field involves development of algorithms that can understand visual information, such as images and videos, like how humans use their eyes. Writing an explicit set of rules for computers to understand any image or video is, again, nearly impossible.

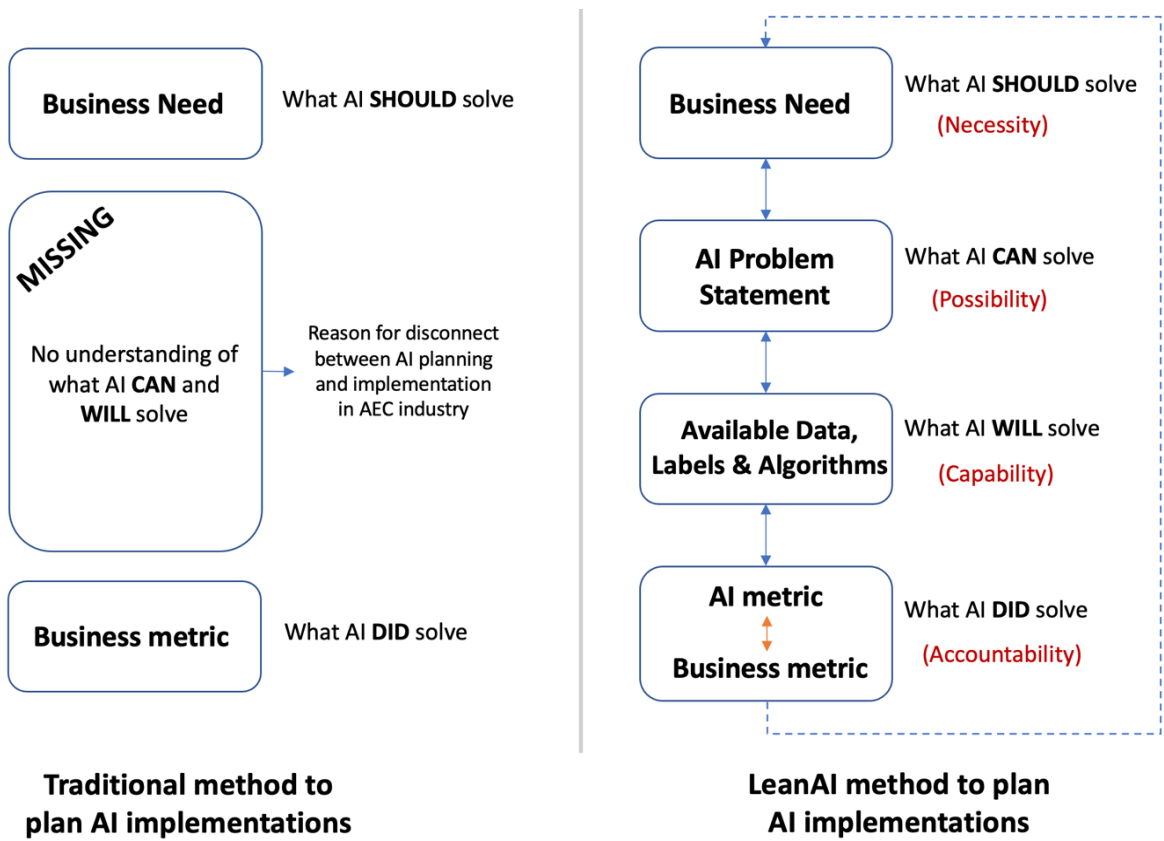

Figure 1 Comparison of traditional and LeanAI method to plan AI implementations

Owing to these exciting possibilities, AI has gained significant interest in the AEC industry [23]. Many pilot AI applications in the AEC industry have already been built to better manage planning of site logistics [24], plan safety on construction sites [25], monitor progress and productivity improvements [26], and plan building operations and maintenance [27].

Although the above pilot applications are encouraging, AI adoption is still in its very early stages in the AEC industry [28]. Limited studies exist that study the practical problems faced by practitioners [8], [29] while planning and deploying AI implementations. One possible reason for this might be that actual AI implementation span over several months or years, making it difficult to conduct a longitudinal study.

Furthermore, the limited interest of practitioners to collaborate and give access to information over such a prolonged period further complicates the issue. The authors of the paper are humbled and grateful for the positive reception they have received from practitioners, which allowed them to create the LeanAI method, which will be discussed in the following section.

## 4 LeanAI Method: An Introduction

When planning AI implementations, AEC practitioners often start by identifying the business needs that AI should address. These business needs, such as reducing maintenance costs for highways or speeding up projects by 20%, are the necessities that the project needs to address and thus drive the AI implementation and ultimately guide what the AI should aim to solve.

However, AI cannot directly solve these business needs. The AI problems need to be framed in an entirely different way as it concerns itself with learning a mathematical function from data. Bridging this gap between high-level business objectives and detailed properties of the implementation is often the key challenge to creating a successful AI project [30].

In our study, we found that AEC practitioners often use a traditional method of planning AI implementations that does not clearly link high-level business objectives with detailed implementation properties (as shown on the left in Figure 1). Practitioners fail to distinguish between what AI should solve, what it can solve, and what it will solve, treating these categories as if they are interchangeable. While AEC practitioners may have a clear understanding of the pressing business needs and may wish that AI should solve certain problems, it is not always possible for AI to address these needs directly. Instead, AI can only solve a small part of the problem when properly formulated, with a direct or indirect impact on the business need. Additionally, just because AI has the capability to solve a problem does not guarantee that it will be able to do so for the practitioner's use case, as it depends on the data and labels available with them.

In fact, over 70% of the projects we observed either did not define a clear problem statement for AI or were unclear about it (e.g., only defining broad business objectives). A similar phenomenon is noted in [31], which states that many AI projects fail because they do not specify the exact problem that AI can solve and expect it to do everything. [30] notes that not defining the problem or success metric for an AI project is a sure way to waste time and money.

The LeanAI method (as shown on the right in Figure 1) clearly delineates what AI should solve, what it can solve, and what it will solve. This approach thus forces practitioners to clearly articulate these components clearly in the planning process itself by involving the relevant stakeholders (e.g., IT team who know about data and labels). Our data suggests that this approach, at least in part, ensures more robust planning of AI implementations.

The following paragraphs will describe the individual elements of the LeanAI method and provide steps for practitioners on how to use the method. We will also illustrate the use of the method through an example case study from our dataset.

**Business need (Necessity that AI should solve):** Defining the business need is an essential first step in planning AI implementations for AEC practitioners. It involves identifying areas where AI should make improvements in the current way of working. Examples of common business needs and necessities include reducing project costs, increasing safety on job sites, and speeding up construction schedules. Specifying a clear business need helps to guide the AI implementation and ensures that the correct problem is being addressed. It also helps to clarify the scope of the project and gain support from top management.

**AI Problem statement (Possibility that AI can solve):** The goal of AI is to address the defined business need, but it cannot do so directly. Instead, the problem statement must be framed in a way that AI can solve it, and at the same time the formulated problem statement, at least partially, addresses the business need. For example, to address the business need of increasing safety on site, one of the problems that AI can potentially solve is to detect that safe distances are maintained between the crane and the workers using the camera image of the site. Another problem statement addressing the same business need can be to detect that workers are wearing protective gear on site. Based on the availability of current AI algorithms, each of these problem statements can be potentially solved by AI. Now, it is up to the practitioners to decide which solution will have the greatest impact for their use case, or to come up with a different AI problem statement altogether.

**Available Data, Labels, and Algorithms (Capability that AI will require):** Many contemporary AI and ML algorithms rely heavily on a large amount of data and labels for optimal performance. Therefore, just because a problem statement has been successfully addressed by AI in one context, it does not guarantee that it will be solved in another context as well. The ultimate determinant of what AI will solve will depend on the availability of suitable data, labels, and the expertise required to develop the algorithm.

**Metrics [AI metric and Business metric] (Accountability of what AI did):** Metrics are the way to measure how well AI did. This measurement process helps to ensure that the AI project stays on track and also provides feedback for continuous improvement in subsequent iterations of AI prototype development. As we have an AI specific problem statement (which is mathematical in nature) and a business need, we need two metrics, one for each, and a way to connect them. That is what we refer to an AI metric and a business metric.

The AI metric measures the performance of the AI algorithm on the problem statement, typically through mathematical measurements such as accuracy, F1 score, etc. which may or may not have a direct correlation to the business. On the other hand, the business metric directly measures the impact on the defined business need. Establishing the relationship between both metrics is a crucial step for practitioners to holistically evaluate AI. For example, in certain cases, 70% accuracy for AI algorithms might result in 25% cost savings and in other cases might not provide any substantial value to the business.

### 4.1 Steps for practitioners to use LeanAI method

While there is no "right" way to use the LeanAI method, we suggest the following steps as a starting point,

especially for the first-time users:

*Step-1: Start by defining the business need:* We have observed in our case studies that practitioners who do not think of the "bigger picture" (aka business need), often fail to get attention from the top management. This, ultimately, results in insufficient resources and lack of traction within the company even if the project is somewhat successful. Therefore, defining the business need is an essential first step. While defining the business need, we suggest practitioners to be as specific and precise as possible, because these business needs are the foundation on which the problem statements would build on later. For example, stating the need as "improving project performance" is not specific enough. It is important to clarify if the goal is to reduce cost, reduce time, improve safety, enhance worker's health, or something else. A more specific and precise business need would be "reduce the cost of the project" or, even better, "reduce the maintenance cost of the project."

*Step-2: Formulate multiple problem statements to address the business need:* Once the specific and precise business need is identified, practitioners should brainstorm multiple ways in which AI can potentially address it. This would involve creating well-defined problem statements that AI can solve, and which will also fulfill the business need. For this step, it is, therefore necessary to involve two types of people: (1) those who are familiar with recent developments in AI and understand what it can and cannot do, and (2) those who know operational details of the business to connect where the capabilities of AI can be utilized.

*Step-3: Evaluating if AI will be able to solve the formulated problem statements:* In the next step, practitioners must determine if they possess the necessary data, labels, and expertise to construct an AI algorithm that can solve the formulated problem. To accomplish this, it is crucial to involve someone with experience in building AI algorithms as the data requirements for these algorithms can, sometimes, be substantial and may not be feasible for many companies. Therefore, it is crucial to obtain a realistic estimate of the data requirements and the performance of the AI algorithm from an expert, to ensure the feasibility of the project.

*Step-4: Defining the metrics to track how well AI did:* Next, practitioners need to define both the AI metric and the business metric and create a link between them. It is important to obtain input from AI experts at this stage to ensure that the metrics and targets set are realistic and achievable, given the data and algorithms at hand. For example, in one case study, the company required an AI algorithm to achieve an accuracy of over 99% for it to have any value to the business. This unrealistic expectation from AI ultimately led to the failure of the AI project, which could have been avoided if the planning process had been done correctly.

*Step-5: Keep iterating:* Creating a successful AI algorithm is not a one-shot task. Practitioners need to continuously iterate and improve depending on how well the AI algorithm is performing on the AI and business metrics. This may sometimes involve revisiting and revising the business need and problem statement either partially or completely.

## 4.2 Example case study demonstrating the use of LeanAI method

In this section, we describe one of our case studies that required practitioners to use AI to improve highway maintenance. We will first explain how practitioners planned this AI implementation using the traditional method. We then illustrate how the LeanAI method helped them improve their planning.

Using traditional planning, we observed practitioners often defined broad and general goals on how AI can help improve highway maintenance, such as reducing the costs, decreasing rework, and increasing daily traffic capacity on highways. While these goals may be areas where AI should have an impact, they were not specific enough to guide the development team in determining how AI can be implemented to achieve them. This lack of specificity led to disconnects during the implementation phase as the AI development team struggled to understand how to apply AI to solve these problems. For example, one practitioner reported that despite top management's push to implement AI on their highway maintenance projects, they have been struggling to find a use case for it. Another practitioner reported that his company's AI project has been stalled for a few months due to a lack of labeled data which they did not account for in the planning phase.

In the LeanAI method, practitioners begin by identifying the business need, such as reducing costs. The next step is to identify the problem statement that AI can solve to address the business need. For example, in the context of highway maintenance, problem statements from practitioners included using AI to automate the detection of cracks on highways, predict crack growth and propagation, and predict traffic growth on the highway. These problem statements all aim to reduce costs by reducing the number of workers needed for maintenance or better planning for maintenance.

Although AI can potentially address all these problem statements, whether it will be able to do so or not will depend on the availability of data, labels, and algorithms. For example, when practitioners evaluated the requirement of data and labels for the various formulated problem statements, they realized the lack of sufficient historical data for predicting crack propagation and growth. Therefore, they had to restrict themselves to either detecting cracks or predicting traffic growth on highways.

Practitioners then established the metrics to track AI and business performance. They determined that "accuracy" would be used to evaluate the AI's performance and the "amount of dollars saved in one year" would measure the business performance. However, upon further analysis, they found that even if the AI achieved the highest level of accuracy in predicting traffic, the financial savings would not justify the AI project. As a result, they decided to proceed with the problem statement of detecting cracks on highways and abandoned the predicting traffic problem statement.

In summary, using the LeanAI method in this case study, practitioners were able to clearly identify a use case for AI that had significant business impact and could be addressed with the available resources within the company, something that was difficult to achieve with the traditional method of planning AI implementations.

## 5 Validation of LeanAI method's usefulness

30 practitioners were divided into ten groups, each consisting of three members. These groups were given 90 minutes to plan an AI implementation project that they were either currently working on or planning to work on in the future, using the LeanAI method. The session ended with a collective reflection session, during which qualitative and quantitative feedback was gathered on the usefulness of the method.

Of the ten groups, seven found the LeanAI method to be very useful, while the remaining three found it somewhat useful. None of the groups found the method to be not useful at all. Practitioners particularly appreciated the simplicity of the method and its practicality, as well as the emphasis it placed on aligning all of the project's elements (business need, problem statement, data, and metrics) together. Participants stated that while each of these elements could be formulated individually, the real challenge was in aligning them, and the LeanAI method helped them achieve this.

Four out of the ten groups expressed interest in conducting more detailed workshops within their companies, indicating that they found the LeanAI method to be very useful.

## 6 Conclusion

The use of AI in the AEC industry presents great opportunity for automation. However, for a wide scale adoption of AI, it is crucial that practitioners have a clear understanding of what AI should solve, what it can solve, and what it will solve. The LeanAI method provides a starting point to address this challenge by encouraging practitioners to involve relevant stakeholders in the early stages of planning, ensuring that AI implementation is aligned with the business needs and goals. This approach can help to bridge the gap between planning and implementation, increasing the chances of success for AI projects in the AEC industry. Future research can further validate the robustness and usefulness of the LeanAI method in AEC and other industries.